\documentclass[aps,prb,reprint,floatfix,amsfonts,amsmath,amssymb,superscriptaddress,showpacs]{revtex4-2}

\usepackage{graphicx,dcolumn,bm,xcolor}

\begin{document}

\title{Direct Electrical Probing of Anomalous Nernst Conductivity}

\author{Weinan Zhou}
\email{ZHOU.Weinan@nims.go.jp}
\affiliation{International Center for Young Scientists, National Institute for Materials Science, Tsukuba 305-0047, Japan}

\author{Asuka Miura}
\altaffiliation[Present address: ]{Integrated Research for Energy and Environment Advanced Technology, Kyushu Institute of Technology, Fukuoka 804-8550, Japan}
\affiliation{Research Center for Magnetic and Spintronic Materials, National Institute for Materials Science, Tsukuba 305-0047, Japan}

\author{Yuya Sakuraba}
\affiliation{Research Center for Magnetic and Spintronic Materials, National Institute for Materials Science, Tsukuba 305-0047, Japan}

\author{Ken-ichi Uchida}
\email{UCHIDA.Kenichi@nims.go.jp}
\affiliation{Research Center for Magnetic and Spintronic Materials, National Institute for Materials Science, Tsukuba 305-0047, Japan}
\affiliation{Institute for Materials Research, Tohoku University, Sendai 980-8577, Japan}

\begin{abstract}
Despite the usefulness of the anomalous Nernst conductivity ($\alpha^{A}_{xy}$) for studying electronic band structures and exploring magnetic materials with large transverse thermopower, there has not been a straightforward way to obtain $\alpha^{A}_{xy}$ in the experiment.
Here, we propose a simple and versatile method enabling direct electrical probing of $\alpha^{A}_{xy}$, which is realized by creating a closed circuit consisting of a target magnetic material and a non-magnetic conductor.
This method was experimentally demonstrated on a thin film of magnetic Weyl semimetal Co$_{2}$MnGa, where the closed circuit was formed simply by connecting both ends of the Co$_{2}$MnGa film with a Au wire.
A good approximation of $\alpha^{A}_{xy}$ was obtained in a wide temperature range, validating the proposed method and exhibiting its potential for aiding the further development of topological materials science and transverse thermoelectrics.
\end{abstract}

\maketitle

\section{Introduction}

The anomalous Nernst conductivity, \textit{i.e.}, the off-diagonal component of the thermoelectric conductivity tensor ($\alpha^{A}_{xy}$) stemming from magnetic moments, describes an intrinsic material property that directly converts a longitudinal temperature gradient into a transverse electric field in a magnetic material.
It has been shown that $\alpha^{A}_{xy}$ is closely linked to the Berry curvature of the electronic bands; in comparison with the anomalous Hall conductivity, which is determined by all occupied bands, $\alpha^{A}_{xy}$ can be more sensitive to the electronic band structures close to the Fermi level, rendering it a valuable tool to study the topological features of magnetic materials through transport measurements \cite{a1,a2,a3,a4,a5,a6,a7,a8,a9,a10,a11,a12,a13,a14,a15,a16,a17,a18}.
In addition to this rapidly increasing interest from the viewpoint of fundamental physics, $\alpha^{A}_{xy}$ is regarded as a crucial parameter to explain unconventionally large transverse thermoelectric output in some magnetic materials where intrinsic contribution plays a dominant role.
Therefore, exploring magnetic materials with large values of $\alpha^{A}_{xy}$ has become a major strategy for thermoelectric applications \cite{b1,b2,b3,b4,b5}.
Due to the orthogonal relationship between the applied temperature gradient and generated electric field, the transverse thermoelectric generation module can be a simple slab or sheet, where no complicated three-dimensional structures are necessary unlike conventional Seebeck-effect-based modules.
Thus, transverse thermoelectric modules could potentially circumvent the problems of durability, flexibility, and cost that the Seebeck modules encounter \cite{ANE1,HFS1,b4,b5,ANE2}, as well as be exploited for additional functionalities, such as heat flux sensing \cite{HFS1,b5,HFS2,HFS3}.
Despite the significant role of $\alpha^{A}_{xy}$ in topological materials science and transverse thermoelectrics, there has not been a straightforward way to experimentally obtain $\alpha^{A}_{xy}$, and establishing such a method is of great importance.

The conventional experimental method for estimating $\alpha^{A}_{xy}$ consists of the measurements of the anomalous Nernst effect (ANE), anomalous Hall effect (AHE), Seebeck effect (SE), and electrical resistivity of a magnetic material.
The anomalous Nernst coefficient ($S_\mathrm{ANE}$), \textit{i.e.}, the transverse thermopower due to ANE, is expressed as
\begin{equation}
\mathit{S}_\mathrm{ANE}=\mathit{\rho}_{xx}\mathit{\alpha}^{A}_{xy}-\mathit{\rho}_\mathrm{AHE}\mathit{\alpha}_{xx}, \label{eq1}
\end{equation}
where $\rho_{xx}$, $\rho_\mathrm{AHE}$, and $\alpha_{xx}$ are the longitudinal resistivity, anomalous Hall resistivity, and diagonal component of the thermoelectric conductivity tensor, respectively.
The first term on the right-hand side of Eq.~(\ref{eq1}) ($S_\mathrm{I}=\rho_{xx}\alpha^{A}_{xy}$) is regarded as an intrinsic component of ANE, while the second term appears as a consequence of AHE acting on the longitudinal electric field induced by SE, which can be rewritten as $S_\mathrm{II}=-S_\mathrm{M}\mathrm{tan}(\theta_\mathrm{AHE})$ [Fig.~\ref{Fig1}(a)] with $S_\mathrm{M}$ being the Seebeck coefficient of the magnetic material and $\mathrm{tan}(\theta_\mathrm{AHE})=\rho_\mathrm{AHE}/\rho_{xx}$ being the anomalous Hall angle.
As a result, $\alpha^{A}_{xy}$ is obtained by experimentally measuring all four parameters of $\rho_{xx}$, $\rho_\mathrm{AHE}$, $S_\mathrm{SE}$, and $S_\mathrm{ANE}$, then calculating using Eq.~(\ref{eq1}).
Many studies have exploited this conventional method to obtain $\alpha^{A}_{xy}$ of a variety of magnetic materials \cite{a2,a3,a4,a5,a7,a8,a9,a10,a11,a12,a13,a14,a15,a16,a17,a18,b1,b4,b5,HFS1,HFS3}.
However, such a task could be cumbersome, and sometimes challenging to complete, since it requires various experimental techniques and measurement systems.

In this study, we propose a method to directly measure the intrinsic component of ANE of a magnetic material and probe its $\alpha^{A}_{xy}$ with ease.
This method is realized simply by creating a closed circuit consisting of the target magnetic material and a non-magnetic conductor, and then measuring transverse thermopower, as shown in Fig.~\ref{Fig1}(b).
The formation of the closed circuit tunes the boundary conditions for electron transport, resulting in the direct emergence of $\alpha^{A}_{xy}$ reflecting the Berry curvature in the transverse thermopower.
We experimentally demonstrated this method by measuring the temperature dependence of $\alpha^{A}_{xy}$ of a Co$_{2}$MnGa thin film, and compared the result with that obtained using the conventional method.
The proposed method grants easy access to $\alpha^{A}_{xy}$, and could be a useful tool in studying topological features and transverse thermoelectric conversion properties of magnetic materials.

\section{Formulation}

\begin{figure}
\includegraphics{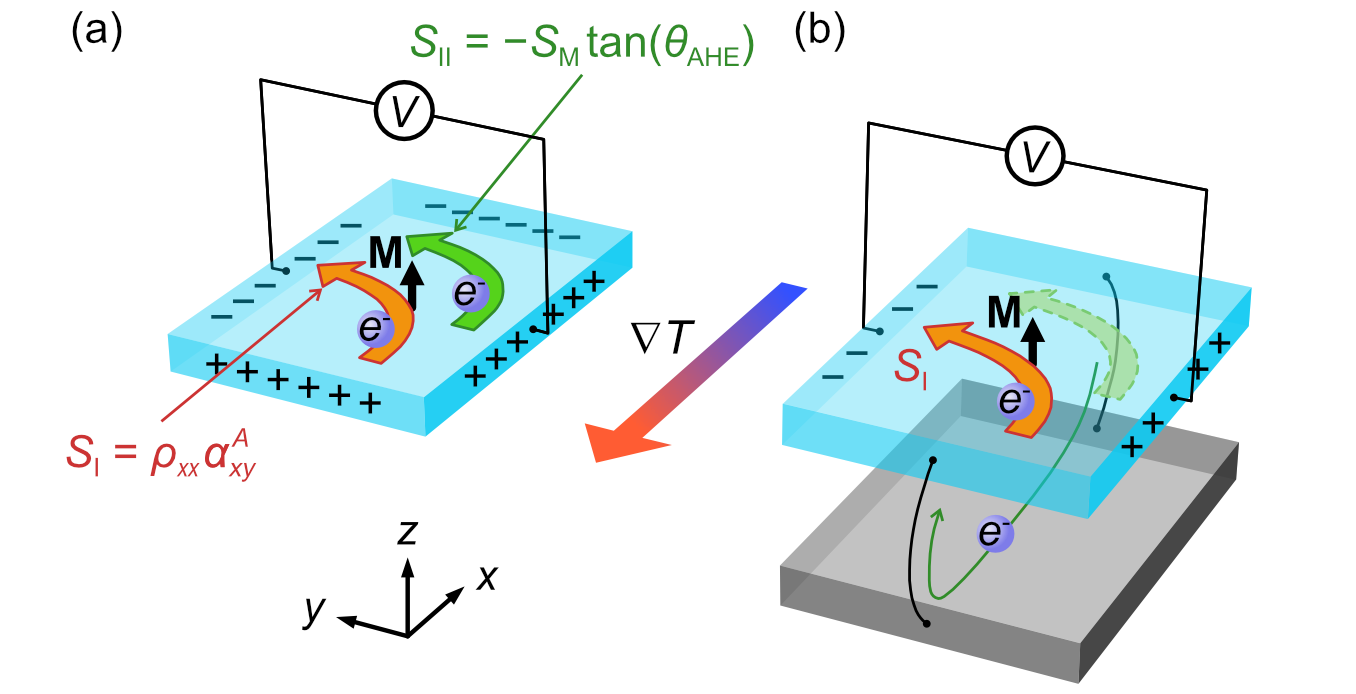}
\caption{\label{Fig1} (a) Schematic illustration of ANE in a magnetic material. The orange and green arrows represent the contribution from the $S_\mathrm{I}$ and $S_\mathrm{II}$ terms of $S_\mathrm{ANE}$, while the black arrow represents the direction of magnetization (\textbf{M}). The $+$ and $-$ symbols indicate the accumulated electric charges due to SE and ANE. (b) Schematic illustration of the closed circuit in which a magnetic material (cyan) is electrically connected to a non-magnetic conductor (gray) at both ends along the direction of the applied temperature gradient ($\nabla{T}$).}
\end{figure}

When a magnetic material is electrically connected to a non-magnetic conductor at both ends along the direction of the applied temperature gradient ($\nabla{T}$), a closed circuit is formed, and its total transverse thermopower measured at the magnetic material ($S^{y}_\mathrm{tot}$) is derived to be \cite{STTG1,STTG2}
\begin{equation}
S^{y}_\mathrm{tot}=S_\mathrm{ANE}-\frac{\rho_\mathrm{AHE}}{\rho_\mathrm{C}/r+\rho_\mathrm{M}}(S_\mathrm{C}-S_\mathrm{M}). \label{eq2}
\end{equation}
Here, $\rho_\mathrm{C(M)}$ and $S_\mathrm{C(M)}$ are the longitudinal resistivity and Seebeck coefficient of the non-magnetic conductor (magnetic material), respectively.
The size ratio $r$ is determined by the geometry of the closed circuit, and in this case, can be expressed as $r=(L_\mathrm{M}/L_\mathrm{C})\times(A_\mathrm{C}/A_\mathrm{M})$, where $L_\mathrm{C(M)}$ is the length of the non-magnetic conductor (magnetic material) along the closed circuit [$x$ axis in Fig.~\ref{Fig1}(b)] and $A_\mathrm{C(M)}$ is the cross-section area of the non-magnetic conductor (magnetic material) perpendicular to the $L_\mathrm{C(M)}$ direction [$yz$ plane in Fig.~\ref{Fig1}(b)].
Previously, thermoelectric materials have been connected to magnetic materials to create closed circuits in order to generate large transverse thermopower \cite{STTG1,STTG3,STTG4}, which is referred to as the Seebeck-driven transverse thermoelectric generation.
However, Eq.~(\ref{eq2}) is still valid when a non-magnetic conductor having small SE is used instead of thermoelectric materials.
If we make $\rho_\mathrm{C}/r\ll\rho_\mathrm{M}$ through small $\rho_\mathrm{C}$, large $r$, or both, the second term on the right-hand side of Eq.~(\ref{eq2}) is reduced to $-S_\mathrm{C}\mathrm{tan}(\theta_\mathrm{AHE})+S_\mathrm{M}\mathrm{tan}(\theta_\mathrm{AHE})$.
Here, $S_\mathrm{M}\mathrm{tan}(\theta_\mathrm{AHE})$ cancels out the $S_\mathrm{II}$ term in $S_\mathrm{ANE}$.
If $\left|S_\mathrm{C}\mathrm{tan}(\theta_\mathrm{AHE})\right|\ll\left|S_\mathrm{I}\right|$, $S^{y}_\mathrm{tot}$ can be used to approximate $S_\mathrm{I}$.
In other words, the magnetic material shunted by connecting to the non-magnetic conductor leads to the disappearance of $S_\mathrm{II}$ [Fig.~\ref{Fig1}(b)].
Then, $\alpha^{A}_{xy}$ can be easily obtained as
\begin{equation}
\alpha^{A}_{xy}\approx\frac{S^{y}_\mathrm{tot}}{\rho_\mathrm{M}}, \label{eq3}
\end{equation}
meaning that the simple shunting process transforms $\alpha^{A}_{xy}$ into a direct experimental observable.
In comparison with the conventional method based on Eq.~(\ref{eq1}), the method proposed here reduces the required parameters for obtaining $\alpha^{A}_{xy}$ from four to two.
If $\rho_\mathrm{M}$ is known, a simple measurement of $S^{y}_\mathrm{tot}$ in the closed circuit enables the direct probing of $\alpha^{A}_{xy}$.

\section{Results and discussion}

\begin{figure}
\includegraphics{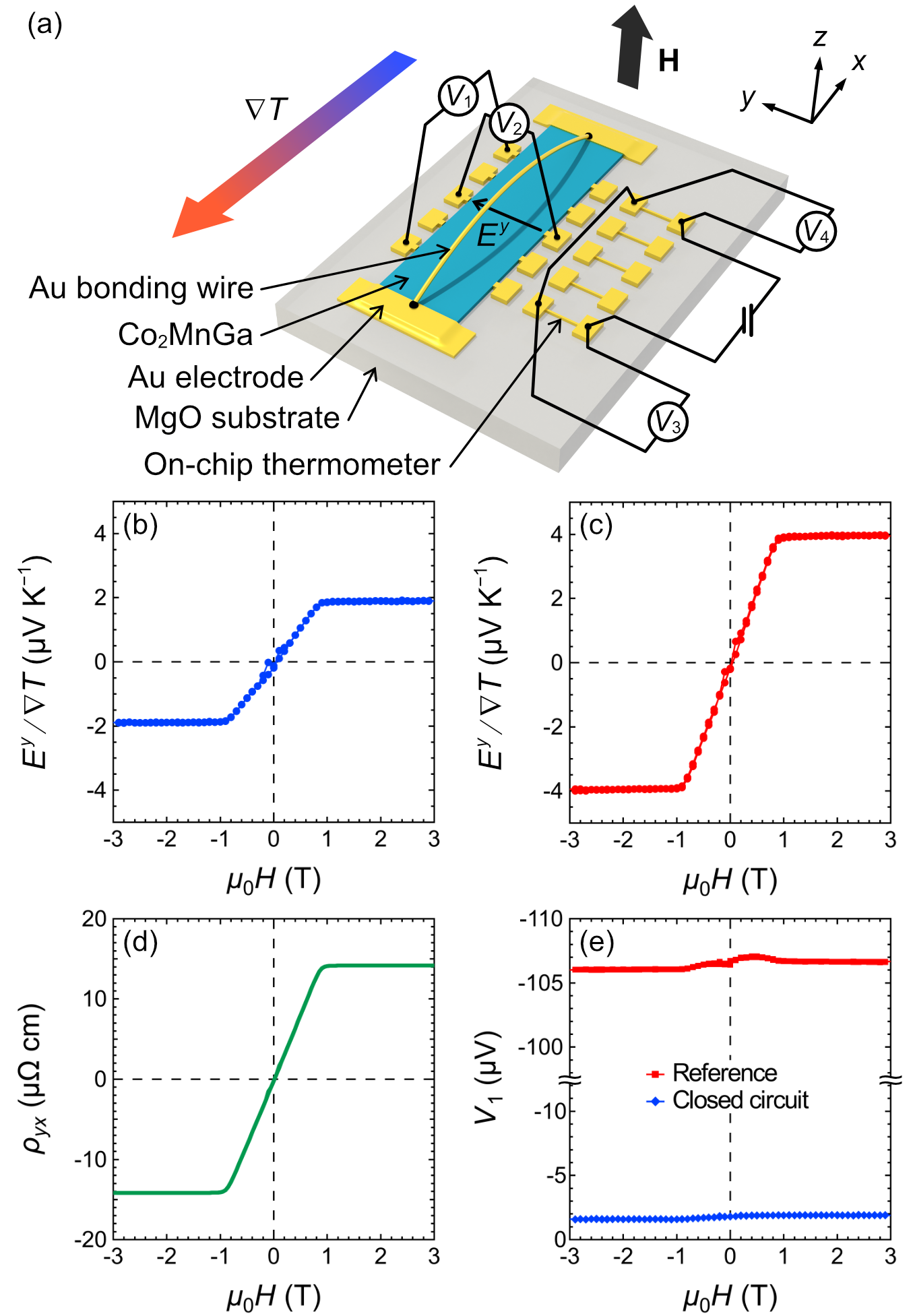}
\caption{\label{Fig2} (a) Schematic illustration of the sample structure and measurement setup for the experimental demonstration of the proposed method to directly probe $\alpha^{A}_{xy}$. $V_{1}$, $V_{2}$, $V_{3}$, and $V_{4}$ represent four nanovoltmeters measuring the longitudinal thermoelectric signal, transverse thermoelectric signal, and resistance of two on-chip thermometers, respectively. (b), (c) $H$ dependence of the transverse electric field ($E^{y}$) divided by $\nabla{T}$ for the closed-circuit sample (b) and the reference sample (c), measured with $T$ of PPMS set at 300 K. (d) $H$ dependence of the transverse resistivity ($\rho_{yx}$) of the reference sample at 300 K, showing AHE of Co$_{2}$MnGa. (e) $H$ dependence of the voltage from $V_{1}$ of the closed-circuit (blue diamond) and reference (red square) samples, measured during the acquirement of the results in (b) and (c), respectively. The magneto-Seebeck effect \cite{MagnetoSeebeck} in Co$_{2}$MnGa was found to be negligibly small.}
\end{figure}

We experimentally demonstrated the proposed method using a Co$_{2}$MnGa thin film.
We chose Co$_{2}$MnGa because it is known as a magnetic Weyl semimetal having substantial $S_\mathrm{I}$ and $S_\mathrm{II}$ terms contributing to its large $S_\mathrm{ANE}$ \cite{a7,a11,a14,a15}.
The 26-nm-thick Co$_{2}$MnGa thin film was epitaxially deposited on a single crystal MgO (100) substrate at room temperature by magnetron sputtering, followed by post annealing at 500$^{\circ}$C.
After the sample was cooled down to room temperature, a 2-nm-thick Al capping layer was deposited to prevent oxidization.
In order to study the structure of the Co$_{2}$MnGa thin film, we measured its X-ray diffraction (XRD) patterns with Cu $K_{\alpha}$ radiation, which indicates an epitaxial growth of Co$_{2}$MnGa (see Supplemental Material \cite{SM} for XRD patterns of the thin film).
From these results, we obtained the integrated intensity ratio between the 111 and 444 peaks ($I^{\mathrm{exp}}_{111}/I^{\mathrm{exp}}_{444}$), and between the 002 and 004 peaks ($I^{\mathrm{exp}}_{002}/I^{\mathrm{exp}}_{004}$).
These values were compared to the simulated integrated intensity ratio ($I^{\mathrm{sim}}_{111}/I^{\mathrm{sim}}_{444}$ and $I^{\mathrm{sim}}_{002}/I^{\mathrm{sim}}_{004}$) for a perfectly $L2_{1}$-ordered structure.
The details of the simulation are described in Ref.~\onlinecite{a15}.
The degree of $L2_{1}$ order ($=\sqrt{\frac{I^{\mathrm{exp}}_{111}/I^{\mathrm{exp}}_{444}}{I^{\mathrm{sim}}_{111}/I^{\mathrm{sim}}_{444}}}$) was estimated to be 0.70 for the Co$_{2}$MnGa thin film, while its degree of $B2$ order ($=\sqrt{\frac{I^{\mathrm{exp}}_{002}/I^{\mathrm{exp}}_{004}}{I^{\mathrm{sim}}_{002}/I^{\mathrm{sim}}_{004}}}$) was estimated to be 0.74.
The composition of Co$_{2}$MnGa was determined to be Co$_{45.7}$Mn$_{25.4}$Ga$_{28.9}$ by X-ray fluorescence spectroscopy.
Then, we patterned the Co$_{2}$MnGa film into a 2-mm-wide and 8-mm-long Hall bar structure using photolithography and Ar ion milling, followed by the formation of Ta (2 nm) / Au (50 nm) electrodes and on-chip thermometers through a lift-off process.
The on-chip thermometers were placed at the positions corresponding to the electrodes of the Hall bar along the $x$ axis, as shown in Fig.~\ref{Fig2}(a).
In order to create the closed circuit, we simply connected both ends of the Co$_{2}$MnGa film along the $x$ axis with a 30-$\mu$m-diameter Au bonding wire.
Here, the Co$_{2}$MnGa is the magnetic material under study, while the Au wire serves as the non-magnetic conductor.
Besides the convenience of use of the Au wire, its small $\rho_\mathrm{C}$ ($=$ 2.3 $\mu\Omega$ cm at room temperature) and small $S_\mathrm{C}$ (between 0.7 and 2.0 $\mu$V K$^{-1}$ for the $T$ range measured in this study \cite{Au}) make it a good non-magnetic conductor for the proposed method.
To measure the transverse thermopower, we set the sample on a home-made holder having a similar structure to the one used in Ref.~\onlinecite{Holder}, which can generate $\nabla{T}$ along the $x$ axis in the sample.
During the measurement, the holder was placed in a physical property measurement system (PPMS), which was used to control temperature ($T$) and magnetic field ($H$) (see Appendix A for details of thermoelectric measurement setup).
$\rho_\mathrm{M}$ and $\rho_\mathrm{AHE}$ of Co$_{2}$MnGa as a function of $T$ were also obtained by measuring the reference sample, where the Au wire connecting both ends of the Co$_{2}$MnGa film was removed.

Figures~\ref{Fig2}(b) and \ref{Fig2}(c) show the $H$ dependence of the transverse electric field ($E^{y}$) divided by $\nabla{T}$ for the closed-circuit and reference samples, respectively, measured with $T$ of PPMS set at 300 K.
The observed signal of the reference sample showed the $H$-odd dependence and saturation at $\left|\mu_\mathrm{0}H\right|\sim$ 1 T, which is attributed to ANE of Co$_{2}$MnGa in the open circuit condition.
By contrast, the signal of the closed-circuit sample is smaller than that of the reference sample, although the shapes are similar to each other.
The curve in Fig.~\ref{Fig2}(b) also saturates at $\left|\mu_\mathrm{0}H\right|\sim$ 1 T along the $z$ axis, suggesting the transverse thermopower of the closed-circuit sample is determined by the magnetization (\textbf{M}) of Co$_{2}$MnGa as well.
Figure~\ref{Fig2}(d) shows the $H$ dependence of $\rho_{yx}$ of Co$_{2}$MnGa measured at 300 K, where the signal is mostly due to AHE.
The $S^{y}_\mathrm{tot}$, $S_\mathrm{ANE}$, and $\rho_\mathrm{AHE}$ values were evaluated by extrapolating the curves in Figs.~\ref{Fig2}(b)-\ref{Fig2}(d) at high $H$ after the saturation of \textbf{M} down to zero $H$.
Figure~\ref{Fig2}(e) shows the longitudinal thermopower from $V_{1}$ measured at the same time when the results in Figs.~\ref{Fig2}(b) and (c) were obtained.
In case of the reference sample, this voltage was due to SE of the Co$_{2}$MnGa-Au thermocouple (note that similar Au bonding wires were used to connect the electrodes of the sample to the home-made holder), and $S_\mathrm{M}$ can be calculated by dividing the voltage at zero $H$ with the corresponding difference of $T$ then adding $S_\mathrm{C}$ of Au.
On the other hand, the magnitude of the longitudinal thermopower of the closed-circuit sample was dramatically reduced, indicating that SE of Co$_{2}$MnGa was indeed shunted by the connection to the Au wire at both ends.
The results measured with $T$ of PPMS set at different values are shown in Supplemental Material \cite{SM}.

\begin{figure}
\includegraphics{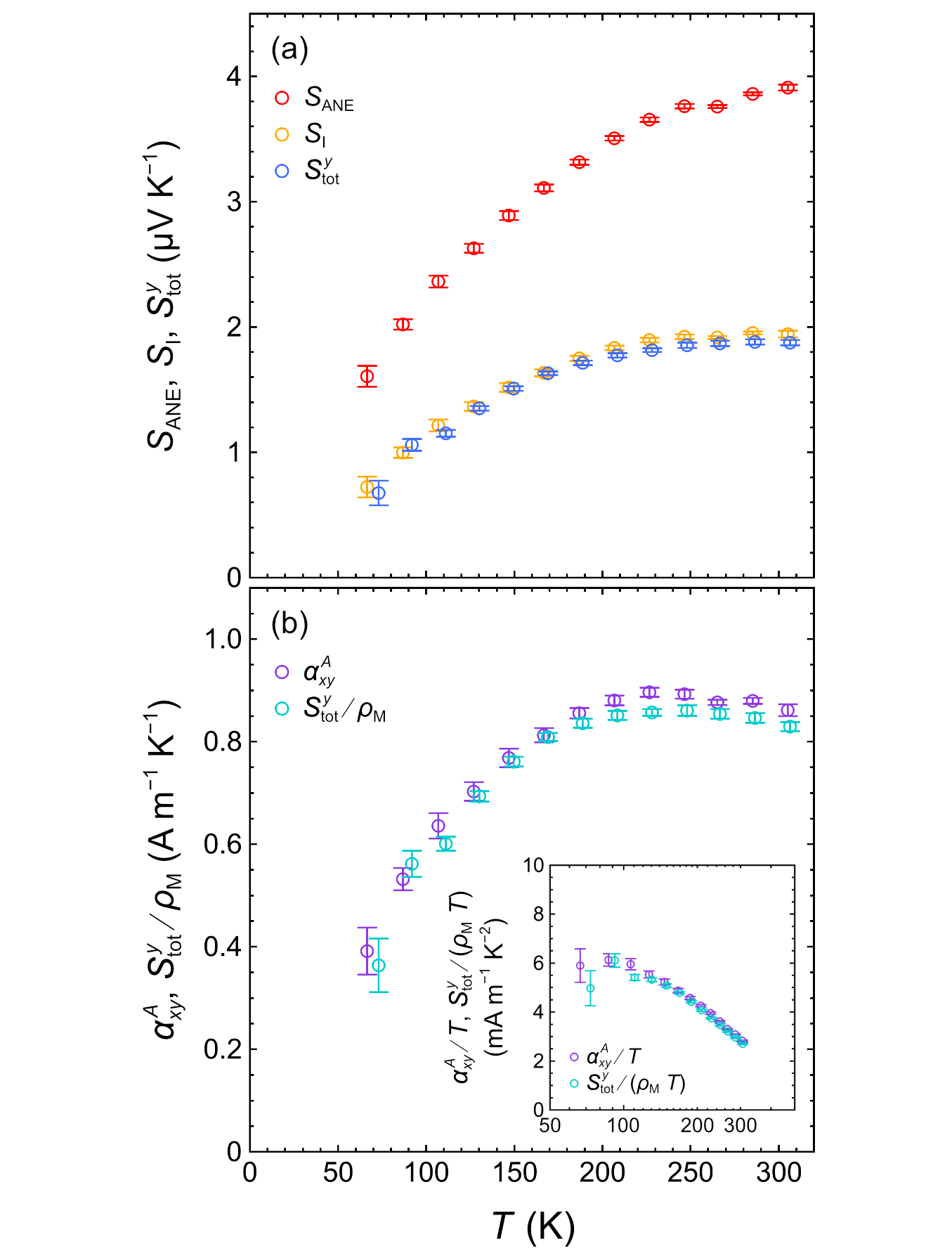}
\caption{\label{Fig3} (a) $S_\mathrm{ANE}$ and $S_\mathrm{I}$ of the reference sample in comparison with $S^{y}_\mathrm{tot}$ of the closed-circuit sample as a funtion of $T$. (b) $\alpha^{A}_{xy}$ obtained using the conventional method and $S^{y}_\mathrm{tot}/\rho_\mathrm{M}$, which approximately corresponds to $\alpha^{A}_{xy}$ through Eq.~(\ref{eq3}). The inset to (b) shows $\alpha^{A}_{xy}/T$ and $S^{y}_\mathrm{tot}/(\rho_\mathrm{M}T)$ plotted as a function of logarithmic $T$.}
\end{figure}

\begin{figure}
\includegraphics{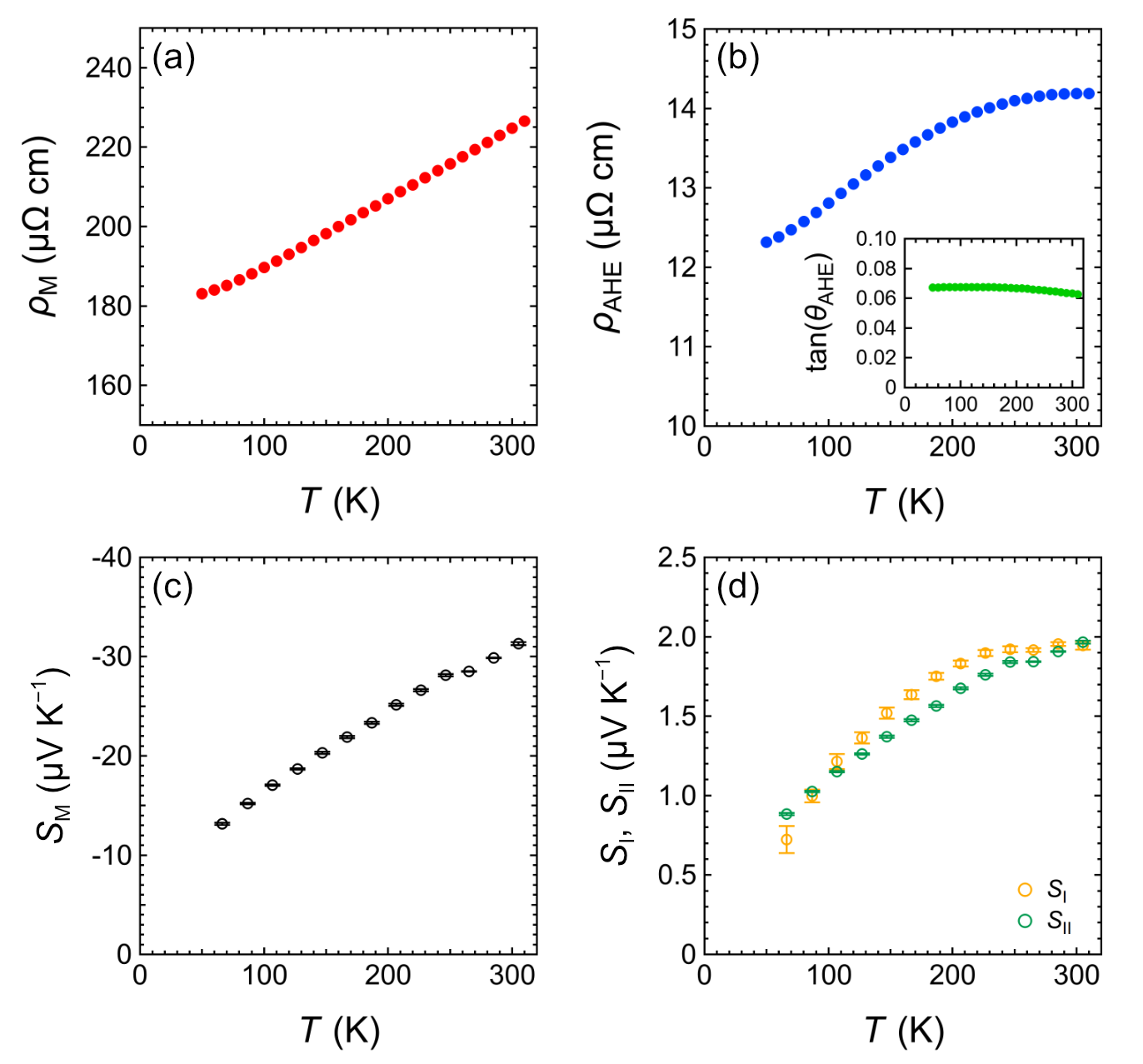}
\caption{\label{Fig4} (a) $\rho_\mathrm{M}$, (b) $\rho_\mathrm{AHE}$, and (c) $S_\mathrm{M}$ of the Co$_{2}$MnGa thin film as a function of $T$ obtained by measuring the reference sample. The inset to (b) shows $\mathrm{tan}(\theta_\mathrm{AHE})$ as a function of $T$. (d) The $S_\mathrm{I}$ and $S_\mathrm{II}$ terms of the Co$_{2}$MnGa thin film as a function of $T$ calculated using the conventional method.}
\end{figure}

By applying Eq.~(\ref{eq3}) to the experimental results of the closed-circuit sample, we were able to probe $\alpha^{A}_{xy}$ of Co$_{2}$MnGa with ease.
The values obtained using the proposed method and the conventional method are compared in Fig.~\ref{Fig3}.
Here, the data points are plotted at $T$ obtained during the measurement of transverse thermopower.
The corresponding $\rho_\mathrm{M}$ and $\rho_\mathrm{AHE}$ of Co$_{2}$MnGa at $T$ were calculated using data points just above and below $T$ with the assumption of linear change with $T$.
Figure~\ref{Fig4} summarizes the $T$ dependence of various parameters of Co$_{2}$MnGa obtained by measuring the reference sample.
These results are necessary to obtain $\alpha^{A}_{xy}$ using the conventional method.
As one can see in Fig.~\ref{Fig3}(a), $S^{y}_\mathrm{tot}$ approximates $S_\mathrm{I}$ nicely and shows the same $T$ dependence, with the values being slightly smaller than $S_\mathrm{I}$.
The same can be said for the comparison between $\alpha^{A}_{xy}$ and $S^{y}_\mathrm{tot}/\rho_\mathrm{M}$.
A logarithmic $T$ dependence of $\alpha^{A}_{xy}/T$ has been considered strong evidence for Weyl magnet Co$_{2}$MnGa \cite{a7}, which can be clearly seen in the inset to Fig.~\ref{Fig3}(b) for both $\alpha^{A}_{xy}/T$ and $S^{y}_\mathrm{tot}/(\rho_\mathrm{M}T)$ in the same $T$ range above $\sim150$ K.
The comparisons thus demonstrate the usefulness of the proposed method in a wide $T$ range.

\begin{figure}
\includegraphics{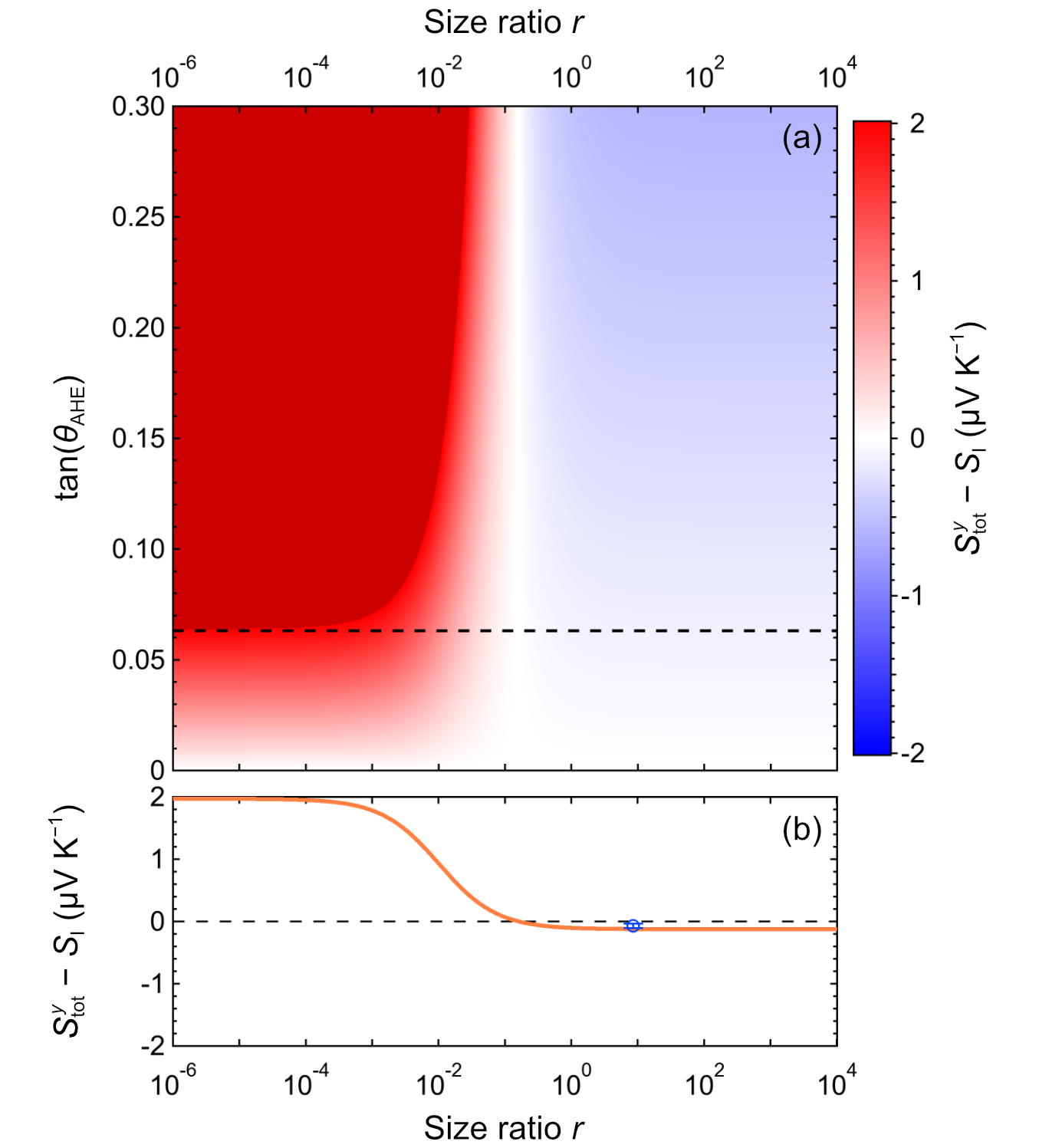}
\caption{\label{Fig5} (a) Calculated $S^{y}_\mathrm{tot}-S_\mathrm{I}$ in the parameter space of size ratio $r$ and $\mathrm{tan}(\theta_\mathrm{AHE})$, with other parameters obtained in the experiment at $T=$ 305 K. Dark red means that $S^{y}_\mathrm{tot}-S_\mathrm{I}>2$ $\mu$V K$^{-1}$. (b) $S^{y}_\mathrm{tot}-S_\mathrm{I}$ as a function of $r$ using $\mathrm{tan}(\theta_\mathrm{AHE})$ of Co$_{2}$MnGa measured in the experiment, which corresponds to the black dashed line in (a). The blue data point represents $S^{y}_\mathrm{tot}-S_\mathrm{I}$ from the experimental results.}
\end{figure}

In order to understand the small difference in results between the two methods and gain insight into the conditions to have a good approximation using the proposed method, we focused on transverse thermopower and calculated $S^{y}_\mathrm{tot}-S_\mathrm{I}$ in the parameter space of the size ratio $r$ and $\mathrm{tan}(\theta_\mathrm{AHE})$, as shown in Fig.~\ref{Fig5}(a).
Here, we use the measured parameters of Co$_{2}$MnGa and $S_\mathrm{C}$ of Au at $T=305$ K for the calculation, except $\rho_\mathrm{AHE}$ (consequently $\mathrm{tan}(\theta_\mathrm{AHE})$) being a varying parameter.
Thus, $S^{y}_\mathrm{tot}-S_\mathrm{I}$ is equivalent to $(-S_\mathrm{M}\mathrm{tan}(\theta_\mathrm{AHE}))-\frac{\rho_\mathrm{AHE}}{\rho_\mathrm{C}/r+\rho_\mathrm{M}}(S_\mathrm{C}-S_\mathrm{M})$ based on Eqs.~(\ref{eq1}) and (\ref{eq2}).
White area in Fig.~\ref{Fig5}(a) represents $S^{y}_\mathrm{tot}-S_\mathrm{I}=0$, while red (blue) represents $S^{y}_\mathrm{tot}$ is larger (smaller) than $S_\mathrm{I}$.
The black dashed line indicates $\mathrm{tan}(\theta_\mathrm{AHE})$ of Co$_{2}$MnGa measured in the experiment, with the curve displayed in Fig.~\ref{Fig5}(b).
$S^{y}_\mathrm{tot}-S_\mathrm{I}$ from the experimental results at $T=$ 305 K is shown as a blue data point in Fig.~\ref{Fig5}(b), where a good agreement between experiment and calculation can be seen.
When $r$ is small, $S^{y}_\mathrm{tot}$ approaches towards $S_\mathrm{ANE}$, and $S^{y}_\mathrm{tot}-S_\mathrm{I}$ approaches towards $S_\mathrm{II}$ (note that $S_\mathrm{II}$ increases as $\mathrm{tan}(\theta_\mathrm{AHE})$ being artificially increased).
As $r$ increases to $>1$ in Fig.~\ref{Fig5}(a), $S^{y}_\mathrm{tot}-S_\mathrm{I}$ decreases and converges to $-S_\mathrm{C}\mathrm{tan}(\theta_\mathrm{AHE})$.
For the majority of magnetic materials having $\mathrm{tan}(\theta_\mathrm{AHE})<0.1$ \cite{AHA}, the value of this difference would be on the scale of $-0.1$ $\mu$V K$^{-1}$.
If the measured $S^{y}_\mathrm{tot}$ is much larger than this value, such as Co$_{2}$MnGa, the proposed method can realize a good approximation of $\alpha^{A}_{xy}$.
If $S^{y}_\mathrm{tot}$ is close to this value, the proposed method is not able to properly evaluate $\alpha^{A}_{xy}$, but it can still serve as an indication of small $S_\mathrm{I}$ and $\alpha^{A}_{xy}$.
Meanwhile, if the target material has an exceptionally large $\mathrm{tan}(\theta_\mathrm{AHE})$, the difference would increase proportionally, although this difference can be further reduced by using a non-magnetic conductor with smaller $S_\mathrm{C}$.
In contrast, the change of $S^{y}_\mathrm{tot}-S_\mathrm{I}$ along the horizontal axis in Fig.~\ref{Fig5}(a) is much more prominent, suggesting that a proper $r$ is important for the proposed method.
Here, $r>$ 1 would be sufficient for $S^{y}_\mathrm{tot}-S_\mathrm{I}$ to reach $-S_\mathrm{C}\mathrm{tan}(\theta_\mathrm{AHE})$, which corresponds to $\rho_\mathrm{C}/r:\rho_\mathrm{M}<1:100$.

As shown above, the proposed method can be easily implemented in the experiment to directly access $S_\mathrm{I}$ of a magnetic material through the measurement of $S^{y}_\mathrm{tot}$, and probe its $\alpha^{A}_{xy}$ by knowing only $\rho_\mathrm{M}$, in contrast to four different parameters required by the conventional method.
This could reduce the number of measurements and ease the requirement on the experimental setup, despite the fact that the measuring process of $S^{y}_\mathrm{tot}$ (applying $\nabla{T}$ to the sample and varying $H$) is the same as that of $S_\mathrm{ANE}$.
The accurate evaluation of $S_\mathrm{M}$ is often difficult due to parasitic thermoelectric signals from electrical wires connected to the sample, $S_\mathrm{M}$ no longer being necessary in the proposed method could be especially beneficial to probe $\alpha^{A}_{xy}$.
Fewer parameters would mean fewer errors propagating to $\alpha^{A}_{xy}$, which could improve the reproductivity and reliability of the results.
These advantages of the proposed method are especially beneficial for high-throughput materials research to determine if a candidate material has large $\alpha^{A}_{xy}$.
In addition, using the first-principles calculations to obtain the Berry curvature and derive $\alpha^{A}_{xy}$ has been popularized in recent years and plays an important role in exploiting and predicting materials with valuable properties.
The proposed method could make $\alpha^{A}_{xy}$ a direct observable in the experiment, thereby enabling fast and straightforward comparison with the theory and promoting further understanding of the matter.
It is worth mentioning that although the experimental demonstration was done on a magnetic thin film, the proposed method should also be applicable to study bulk materials, as long as the required condition of $\rho_\mathrm{C}/r\ll\rho_\mathrm{M}$ is met.

\section{Conclusion}

We have proposed a method to directly probe $\alpha^{A}_{xy}$ of a magnetic material, which is realized simply by connecting both ends of the magnetic material along the direction of $\nabla{T}$ with a non-magnetic conductor to create a closed circuit.
$S^{y}_\mathrm{tot}$ of the closed circuit approximates the $S_\mathrm{I}$ term of the magnetic material, and $\alpha^{A}_{xy}$ can be easily obtained from $S^{y}_\mathrm{tot}$ and $\rho_\mathrm{M}$, in contrast to four different parameters required in the conventional method.
The proposed method was experimentally demonstrated to probe the $T$ dependence of $\alpha^{A}_{xy}$ of a Co$_{2}$MnGa thin film.
The closed circuit was easily realized using an Au wire, and a good approximation was obtained for both $S_\mathrm{I}$ and $\alpha^{A}_{xy}$ in the entire $T$ range between 70 and 305 K, validating this method.
Proper conditions to achieve a good approximation are also discussed, providing guides for the utilization of the proposed method.
As the popularity of using $\alpha^{A}_{xy}$ is growing, our finding could become a powerful tool propelling studies of topological materials science and application of transverse thermoelectric phenomena.

\begin{acknowledgments}
The authors thank R. Toyama and T. Hirai for their support in sample preparation and measurement. This work was supported by JST CREST “Creation of Innovative Core Technologies for Nano-enabled Thermal Management” (Grant No. JPMJCR17I1), JST ERATO “Magnetic Thermal Management Materials” (Grant No. JPMJER2201), JSPS KAKENHI Grant-in-Aid for Scientific Research (B) (Grant No. JP21H01608) and Grant-in-Aid for Research Activity Start-up (Grant No. JP22K20494), and NEC Corporation.
\end{acknowledgments}

\renewcommand{\appendixname}{APPENDIX}
\appendix
\section{THERMOELECTRIC MEASUREMENT SETUP}

For the thermoelectric measurements, we used a home-made holder, which was built upon a commercially available multi-function probe and an empty sample puck.
Here, one side of the sample was thermally connected to a Cu block then to the puck that serves as a heat sink; the other side was thermally connected to a heater through another Cu block while insulated from the puck by a bakelite plate.
This structure to generate temperature gradient ($\nabla{T}$) in a thin film sample is similar to the one used in Ref.~\onlinecite{Holder} (see Supplemental Material \cite{SM} for a photo of thermoelectric measurement setup).
In order to measure the electrical signals from the sample, the same Au bonding wires were used to connect the electrodes on the samples to the electrodes on the holder.
These electrical connections then went to the other end of the multifunction probe through Cu wires, and eventually out of the probe and into a connection box to meet with the cables from external electronics.
It is worth mentioning that the electrodes on the holder were thermally anchored to the heat sink.
During thermoelectric measurements, the home-made holder was placed in a physical property measurement system (PPMS; VersaLab made by Quantum Design).
The heater was electrically connected to the pins at the bottom of PPMS.
These connections then went through the cables inside PPMS and eventually fed to an external source meter (KEITHLEY 2401 SourceMeter).
Another source meter of the same type was used to apply 100 $\mu$A to the on-chip thermometers.
Together with two nanovoltmeters ($V_{3}$ and $V_{4}$ in Fig.~\ref{Fig2}(a); KEITHLEY 2182A NANOVOLTMETER), we were able to accurately measure the resistance of on-chip thermometers by the four-terminal method.
Meanwhile, two nanovoltmeters of the same type ($V_{1}$ and $V_{2}$ in Fig.~\ref{Fig2}(a)) were used to measure the longitudinal and transverse thermoelectric signals from the sample.
All the external electronics were controlled by a PC through GPIB cables, while this PC also communicated with another PC that controlled the temperature ($T$) and magnetic field ($H$) of PPMS.
Prior to generate $\nabla{T}$ in the sample, the on-chip thermometers were calibrated by measuring their resistance as a function of $T$ under zero $H$.
Then, we set $T$ of PPMS to a certain value, generated $\nabla{T}$ in the sample, and swept $H$ along the $z$ axis while monitoring the longitudinal and transverse thermoelectric signals from the closed circuit with two nanovoltmeters, $V_{1}$ and $V_{2}$, respectively.
The resistance of the on-chip thermometers was also measured and subsequently converted to $T$; the difference was divided by the distance between the two thermometers (4 mm) to work out the value of $\nabla{T}$, while the average was used as $T$ for the obtained $S^{y}_\mathrm{tot}$.
$T$ of PPMS was varied between 60 and 300 K with 20 K interval.
The same measuring process was carried out on the reference sample, where the Au wire connecting both ends of the Co$_{2}$MnGa film was removed; this is the conventional ANE measurement.

\bibliography{Text_ref}

\end{document}